\documentclass{aa}  

\usepackage{graphicx}
\usepackage{txfonts}
\usepackage[table,xcdraw,dvipsnames]{xcolor}
\usepackage{lipsum}
\usepackage{subcaption} 
\usepackage{lscape} 
\usepackage{placeins}
\usepackage{sidecap}
\sidecaptionvpos{figure}{c}

\newcommand{\teff}{\mbox{$T_{\rm eff}$}\xspace}
\newcommand{\feh}{\mbox{$\rm{[Fe/H]}$}\xspace}

\newcommand{\numax}{\mbox{$\nu_\mathrm{max}$}\xspace}
\newcommand{\deltanu}{\mbox{$\Delta\nu$}\xspace}

\newcommand{\msun}{\mbox{$\mathrm{M}_{\odot}$}\xspace}
\newcommand{\lsun}{\mbox{$\mathrm{L}_{\odot}$}\xspace}
\newcommand{\rsun}{\mbox{$\mathrm{R}_{\odot}$}\xspace}
\newcommand{\numaxsun}{\mbox{$\nu_\mathrm{max,\odot}$}\xspace}
\newcommand{\deltanusun}{\mbox{$\Delta\nu_\mathrm{\odot}$}\xspace}

\newcommand{\nunl}{\mbox{$\nu_{n \ell}$}\xspace}
\newcommand{\rone}{\mbox{$r_{\rm 01}$}\xspace}
\newcommand{\rtwo}{\mbox{$r_{\rm 02}$}\xspace}

\usepackage{hyperref}
\hypersetup{
    colorlinks=true,   
    linkcolor=blue,     
    citecolor=blue,   
    filecolor=magenta,  
    urlcolor=blue     
}

\begin{document}

   \title{Physics-induced impact on the properties of \textit{Kepler}/K2 solar analogs in the light of PLATO's requirements}

   \author{S. Khan\inst{1}\corrauth{skhan@uliege.be}
        \and G. Buldgen\inst{1}
        \and J. Bétrisey\inst{2,3}
        \and R. A. García\inst{4}
        \and S. Mathur\inst{5,6}
        }

   \institute{STAR Institute, Université de Liège, Liège, Belgium
   \and Department of Physics and Astronomy, Uppsala University, Box 516, SE-751 20 Uppsala, Sweden
   \and Institut für Astrophysik, Zürich Universität, Winterthurerstrasse 190, 8057 Zürich, Switzerland
   \and Université Paris-Saclay, Université Paris Cité, CEA, CNRS, AIM, 91191, Gif-sur-Yvette, France
   \and Instituto de Astrofísica de Canarias (IAC), E-38205 La Laguna, Tenerife, Spain
   \and Universidad de La Laguna (ULL), Departamento de Astrofísica, E-38206 La Laguna, Tenerife, Spain}

   \date{Received September 30, 20XX}

  \abstract
   {Solar analogs are stars whose fundamental properties closely match those of the Sun and, as such, are useful laboratories for tests of stellar structure and evolution, particularly in the context of the upcoming PLATO mission.}
   {Our study focuses on six seismic solar analogs observed by \textit{Kepler} and K2, which have been characterised in detail in a former analysis. We investigate the impact of the model physical ingredients on these six targets in order to evaluate how well one can satisfy the PLATO precision requirements for stellar parameters with the data quality that is currently available to us.} 
   {We used as reference values the stellar mass, age, and chemical composition obtained with the Forward and Inverse COmbination (FICO) procedure, based on OPAL opacities and a combination of NACRE II and Solar Fusion II (SFII) nuclear reaction rates. The efficiency of macroscopic transport is calibrated on lithium abundances derived from the analysis of HERMES spectra. From there, we carried out extensive tests of the physical ingredients with the help of a Levenberg-Marquardt local minimisation method using both classical (luminosity, mean density, metallicity, and effective temperature) and seismic constraints (frequency separation ratios). In particular, we investigated the impact induced by the use of OPLIB opacities and SFIII nuclear reaction rates on the stellar fundamental properties, as well as the use of various seismic constraints.}
   {We quantify the potential differences that changes in the modelling input physics may lead to, and compare their extent with the requirements expected for the PLATO mission (15\% in mass, 2\% in radius, and 10\% in age), providing crucial constraints on the accuracy and robustness with respect to systematics of the determination of fundamental properties for PLATO. For our sample of solar analogs, we are able to accurately retrieve stellar masses within 2.5\%. As for stellar ages, we quantify systematics of the order of 10-14\% at most, but there are also very promising cases for which deviations remain within 4\% depending on the stars and constraints considered.}
   {}


   \maketitle
   \nolinenumbers

\section{Introduction}
\label{sec:intro}

In the last decade, the field of asteroseismology has known a space-based revolution with the advent of the space missions Convection, Rotation, and planetary Transits \citep[CoRoT,][]{Baglin2006,Auvergne2009}, \textit{Kepler}/K2 \citep{Borucki2009,Howell2014}, and Transiting Exoplanet Survey Satellite \citep[TESS,][]{Ricker2014}. Another milestone is to be expected early 2027 with the ESA PLAnetary Transits and Oscillations of stars (PLATO) mission, which will observe thousands of solar-like oscillators in a field of view approximately 20 times larger than \textit{Kepler} \citep{Rauer2014,Rauer2025}. The PLATO precision requirements for the stellar fundamental parameters are: 15\% for the mass, 2\% for the radius, and 10\% for the age for a dwarf with an effective temperature of 6000 K.

In light of those requirements, some studies \citep[see, e.g.,][]{Lebreton2014,Betrisey2023,Li2024,Grossmann2025} have been dedicated to the detailed asteroseismic modelling of specific and well-characterised solar-like oscillators, focusing on the best way to make use of asteroseismic constraints (global observables or individual oscillation frequencies) as well as on the impact of varying physical ingredients in the models, such as opacities and nuclear reaction rates, among others. \citet{Nsamba2018,Nsamba2021} studied the impact of atomic diffusion, the solar metallicity mixture, the initial helium abundance, and different surface correction methods on a sample of \textit{Kepler} Legacy solar-type stars. \citet{Betrisey2022} focused on the detailed asteroseismic study of Kepler-93, a solar twin and exoplanet host star. \citet{Moedas2024,Moedas2025} evaluated the combined effect of microscopic transport processes, such as atomic diffusion (with gravitational settling alone or also including radiative accelerations), and of turbulent mixing on \textit{Kepler} Legacy and planet-hosting FGK-type stars. \citet{Buldgen2025} looked into a subsample of the \textit{Kepler} Legacy sample to explore the question of lithium depletion, invoking extra-mixing either in the form of convective overshooting below the envelope or turbulence in radiative layers.

Solar analogs are Sun-like stars whose properties resemble those of the Sun, but without having to match all fundamental properties at once -- unlike solar twins. They are usually contained in the $0.9 \leq M \leq 1.1 \, \msun$ mass range, with no specific constraint on stellar age. This means that they span a wider evolution range with respect to the Sun, with different rotation rates, magnetic activity levels, and chemical compositions. Their study allows for a reconstruction of the time sequence of the Sun. Examples of solar analogs are KIC 10644253 \citep[young and active,][]{Salabert2016}, KIC 8006161 \citep[young and metal-rich,][]{Karoff2018}, and the binary system 16 Cyg A/B \citep[old,][]{Metcalfe2012,Davies2015,Bazot2020,Farnir2020,Buldgen2022}.

More specifically, seismic solar analogs are stars whose global asteroseismic parameters, namely \numax and \deltanu, are found within 10\% of the solar reference values \citep[$\numaxsun=3090\,\rm\mu Hz$ and $\deltanusun=135.1\,\rm\mu Hz$ based on][]{Huber2011}. Despite not necessarily fitting in the definition of solar analogs in a spectroscopic sense, these stars nevertheless offer a precise window into the internal structure of Sun-like stars.

\begin{table*}[]
    \centering
    \caption{Observational constraints for the six \textit{Kepler}/K2 seismic solar analogs analysed in this study.}
    \begin{tabular}{c|cccc}
    \hline\hline
         Target          & Luminosity (\lsun)   & \teff (K)      & \feh (dex)        & A(Li) (dex)    \\
    \hline
         EPIC 206064678  & 0.8025 $\pm$ 0.0061  & 5476 $\pm$ 55  & 0.25 $\pm$ 0.07   & $< 1.0$        \\
         EPIC 206245055  & 1.0119 $\pm$ 0.0067  & 5841 $\pm$ 54  & -0.39 $\pm$ 0.09  & $2.0\pm0.05$   \\
         EPIC 206371648  & 0.8460 $\pm$ 0.0336  & 5573 $\pm$ 43  & -0.26 $\pm$ 0.06  & $< 1.0$        \\
         EPIC 212624487  & 1.3112 $\pm$ 0.0096  & 6038 $\pm$ 27  & -0.16 $\pm$ 0.09  & $2.55\pm0.05$  \\
         EPIC 212708252  & 0.8689 $\pm$ 0.0065  & 5579 $\pm$ 43  & -0.09 $\pm$ 0.07  & $< 0.8$        \\
         KIC 3241581     & 1.1001 $\pm$ 0.0256  & 5730 $\pm$ 50  & 0.25 $\pm$ 0.06   & $< 1.0$        \\ 
    \hline
    \end{tabular}
    \tablefoot{See also Tables 1 and 2 in \citet{Garcia2026}.}
    \label{tab:obs}
\end{table*}

\begin{table*}[]
    \centering
    \caption{Stellar parameters determined from the FICO procedure using standard models.}
    \begin{tabular}{c|c|cccccc}
    \hline\hline
         Target          & Fit                 & Mass (\msun)       & Radius (\rsun)     & Age (Gyr)        & Mean density (g $\rm cm^{-3}$)  & $X_{\rm 0}$  & $Z_{\rm 0}$ \\
    \hline
         EPIC 206064678  & \rone+\;\rtwo (3) & 1.058 $\pm$ 0.021  & 1.004 $\pm$ 0.007  & 5.40 $\pm$ 0.38  & 1.4724 $\pm$ 0.0059             & 0.724        & 0.027       \\
         EPIC 206245055  & \rone+\;\rtwo (3)  & 0.899 $\pm$ 0.020  & 0.943 $\pm$ 0.007  & 6.04 $\pm$ 0.35  & 1.5105 $\pm$ 0.0060             & 0.738        & 0.006       \\
         EPIC 206371648  & \nunl (2)         & 0.947 $\pm$ 0.019  & 0.976 $\pm$ 0.007  & 7.76 $\pm$ 0.55  & 1.4359 $\pm$ 0.0086             & 0.740        & 0.013       \\
         EPIC 212624487  & \rone+\;\rtwo (3) & 1.073 $\pm$ 0.018  & 1.034 $\pm$ 0.006  & 2.03 $\pm$ 0.31  & 1.3704 $\pm$ 0.0055             & 0.740        & 0.011       \\
         EPIC 212708252  & \rone+\;\rtwo (3) & 0.927 $\pm$ 0.010  & 0.981 $\pm$ 0.004  & 9.20 $\pm$ 0.36  & 1.3849 $\pm$ 0.0055             & 0.734        & 0.013       \\
         KIC 3241581     & \nunl (2)          & 1.056 $\pm$ 0.036  & 1.088 $\pm$ 0.012  & 5.60 $\pm$ 0.47  & 1.1610 $\pm$ 0.0093             & 0.687        & 0.027       \\ 
         KIC 3241581     & \nunl (full, 2)   & 1.124 $\pm$ 0.034  & 1.109 $\pm$ 0.012  & 5.07 $\pm$ 0.38  & 1.1610 $\pm$ 0.0093             & 0.725        & 0.026 \\
         \hline
    \end{tabular}
    \tablefoot{The underlying grid of models has been computed with OPAL opacities, a combination of NACRE II and SFII nuclear reaction rates, and no turbulent diffusion \citep[see also Sect. 6.5 and Table 3 in][]{Garcia2026}. In the Fit column, the number in brackets indicates how many FICO steps have been performed to obtain those results.}
    \label{tab:fico}
\end{table*}

Lithium is one of the key chemical elements in the field of stellar evolution. Its fusion temperature is so low ($\sim 2.6 \cdot 10^6 \, \rm K$) that lithium is particularly sensitive to transport processes happening in the interior of stars. Many studies have been dedicated to the analysis of its long-term evolution \citep[see, e.g.,][]{Boesgaard1976,Boesgaard1991,Carlos2019,Carlos2020,Borisov2024a,Borisov2024}, to shed light on the underlying physical processes leading to its depletion \citep[see, e.g.,][]{Baglin1985,Lebreton1987,Montalban1994,Charbonnel1994,Schlattl1999,Montalban2000,Thevenin2017,Baraffe2017,Eggenberger2022,Buldgen2025a} but also to relate it with the presence, or not, of exoplanets around the host stars of concern \citep[see, e.g.,][]{Bouvier2008,Castro2009,Deal2015}. It also plays a crucial role for primordial nucleosynthesis models \citep[see, e.g.,][]{Fields2011,Clara2020,Deal2021}. 

One major difficulty with lithium is that it is significantly burned during the pre-main-sequence phase. And, in the case of a solar-like star, the fusion temperature is very close to the transition temperature between radiative and convective layers. This means that the observed lithium depletion can be reproduced by both the turbulence in the radiative layers and the extra-mixing (overshooting) occurring at the base of the convective envelope. In this context, additional constraints provided by a precise determination of the beryllium abundance usually help disentangling both processes, as its higher fusion temperature makes it impossible to reproduce using only overshooting.

In this work, we analyse six seismic solar analogs observed by \textit{Kepler}/K2, HERMES, and \textit{Gaia}, for which high-quality asteroseismic and spectroscopic data are available. To this end, the first part of the detailed modelling of these targets was carried out in \citet{Garcia2026}, in the form of a multi-pipeline global minimisation. The Forward and Inverse COmbination (FICO) method is used as a reference in the current work for the stellar mass, age, and chemical composition for each of these targets. We build upon these results to run local minimisations with extensive changes to the physical ingredients, notably the opacities and nuclear reaction rates, and the inclusion of additional mixing at the base of the convective envelope to account for lithium depletion, and we quantify their impact on the stellar fundamental parameters. 

\section{Reference sample}
\label{sec:reference}

   We studied six seismic solar analogs that have been presented and studied in detail, using standard stellar models, by \citet{Garcia2026}. Over the years, standard solar models have provided a well-defined framework, with a given combination of physical ingredients, to reproduce the observational properties of the Sun \citep[see, e.g.,][]{Bahcall2001,Serenelli2016,Vinyoles2017}. Hence, in the current paper, standard refers to a fixed set of physical ingredients (here those used by FICO), and non-standard to models computed with changes in, e.g., the opacities, nuclear reaction rates, inclusion of macroscopic transport with respect to the physics originally used in the standard models. Several stellar modelling methods have been used and compared within the analysis of \citet{Garcia2026}, but we focus here on the results obtained with the FICO procedure as a starting point for our analysis with non-standard models. 
   
   The reference values for our six targets were obtained with the FICO method. This procedure combines forward modelling and seismic inversions to mitigate surface effects and improve the estimation of stellar parameters. We refer to \citet{Betrisey2023,Betrisey2026} for a comprehensive description of the methodology as well as of the physical ingredients of the underlying stellar models. We recall here briefly the modelling aspects that are most relevant to our study. The FICO method consists of three steps:
   \begin{itemize}
       \item first, a reference model is obtained by fitting individual oscillation frequencies using the surface correction of \citet{Ball2014}, providing a reliable initial estimate for the stellar mean density;
       \item second, this estimate is refined through a mean-density inversion using the nonlinear extension of the SOLA formalism \citep{Reese2012}, yielding a quasi model-independent constraint;
       \item third and last, the final model is obtained by fitting frequency separation ratios -- together with the inverted mean density and the spectroscopic constraints -- to strongly suppress surface effects \citep{Roxburgh2003,OtiFloranes2005}.
   \end{itemize}

   Model optimisation was carried out with the AIMS software \citep{Rendle2019}, using the grid of stellar models from \citet{Betrisey2023,Betrisey2026}. These models are computed with the Code Liégeois d'Evolution Stellaire \citep[CLES,][]{Scuflaire2008} and have the following physical ingredients: solar abundances from \citet{Asplund2009}, OPAL opacities \citep{Iglesias1996} supplemented with low-temperature opacities from \citet{Ferguson2005}, FreeEOS equation of state \citep{Irwin2012}, and a combination of NACRE II and Solar Fusion II (SFII) nuclear reaction rates \citep{Adelberger2011,Xu2013}. Microscopic diffusion follows \citet{Thoul1994} with screening coefficients from \citet{Paquette1986}. Convection is treated using mixing-length theory with a fixed solar-calibrated value of $\alpha_{\rm MLT}=2.05$. Atmospheres are modelled using the $T(\tau)$ relation from \citet{Vernazza1981}.

\begin{figure}
    \centering
    \includegraphics[width=\hsize]{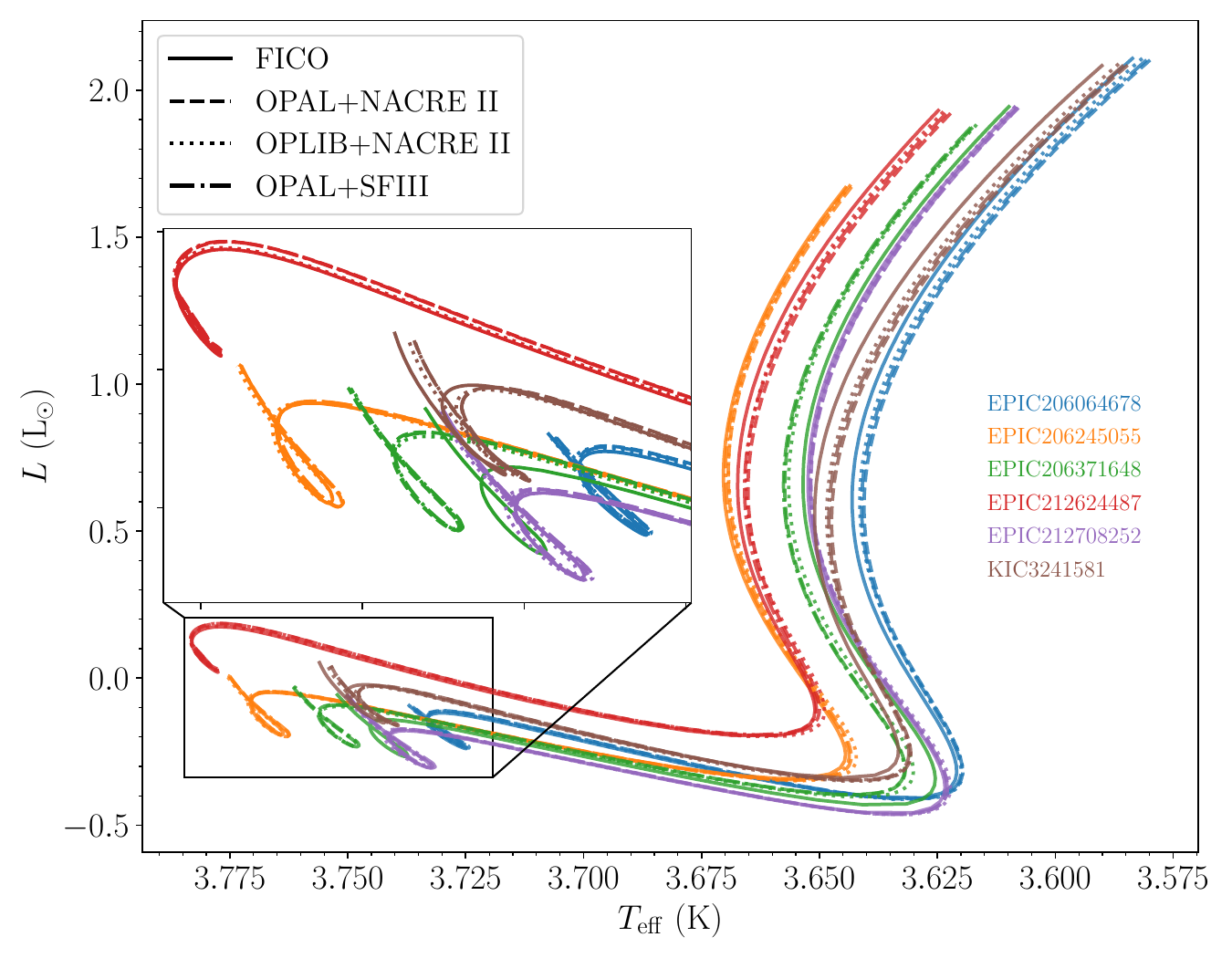}
        \caption{Hertzsprung-Russell diagrams for standard (FICO, solid) and non-standard models (OPAL+NACRE II, dashed; OPLIB+NACRE II, dotted; OPAL+SFIII, dash-dotted lines). Each colour corresponds to a different target, as labelled on the right: blue for EPIC 206064678, orange for EPIC 206245055, green for EPIC 206371648, red for EPIC 212624487, purple for EPIC 212708252, and brown for KIC 3241581. The inset on the left zooms in on the final part of the evolution to help differentiate between the different physics.}
        \label{fig:hrd}
\end{figure}

   The observational constraints and stellar parameters determined with the FICO method, using standard models, for all six seismic solar analogs are given in Tables \ref{tab:obs} and \ref{tab:fico}. The corresponding stellar evolutionary tracks are displayed on the Hertzsprung-Russell diagram in Fig. \ref{fig:hrd} as solid lines.

\begin{figure}
    \centering
    \includegraphics[width=\hsize]{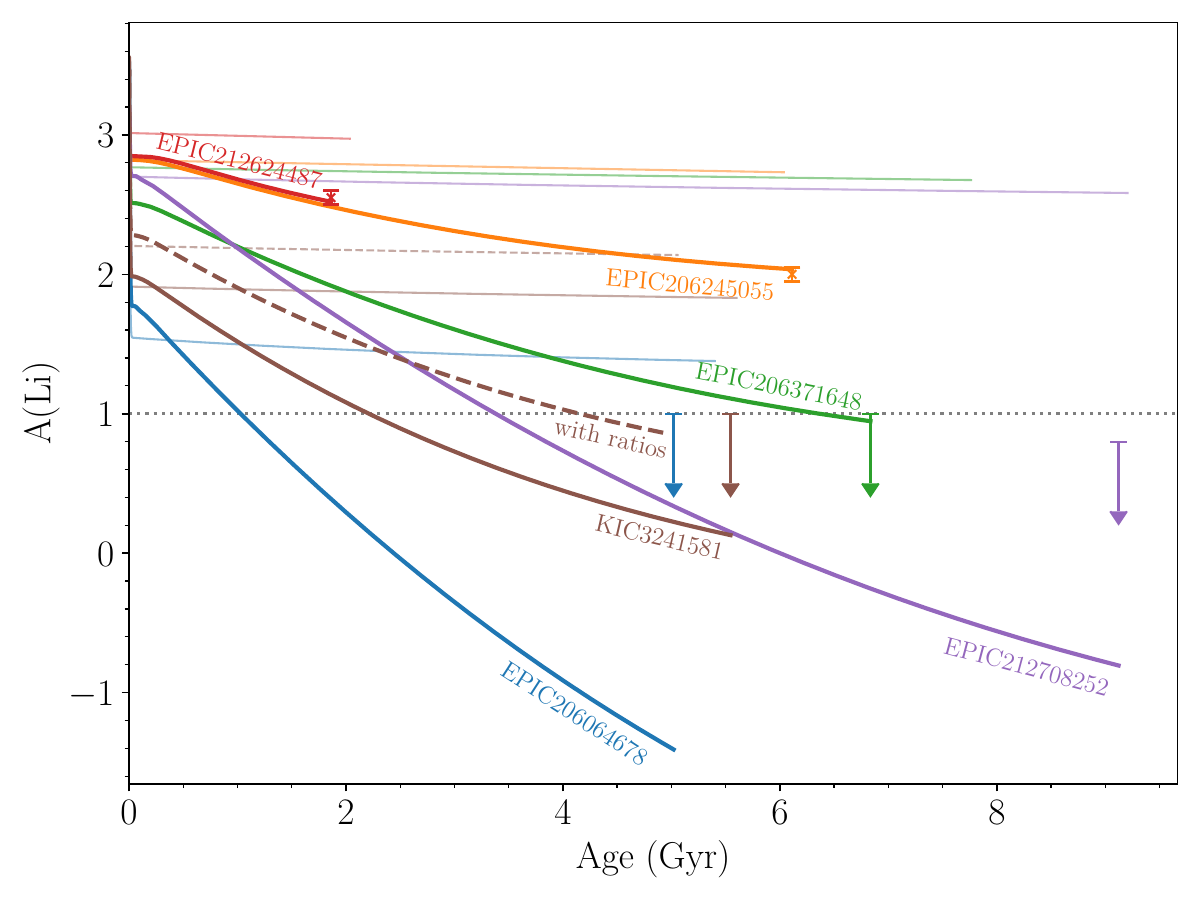}
        \caption{Evolution of the surface lithium abundance A(Li) as a function of age. Precise observational estimates with uncertainties have been obtained for two of our targets (EPIC 206245055 and EPIC 212624487), while for the other four only an upper limit on the value of A(Li) is available. For KIC 3241581, the dashed line corresponds to the minimisation using two ratio estimates on top of the individual frequencies (instead of the frequencies alone). The thin coloured lines show the evolution based on the FICO reference models, which do not include any extra mixing. A grey dotted horizontal line has been drawn at $\rm A(Li) = 1.0$ to guide the eye.}
        \label{fig:agelithium}
\end{figure}

\begin{figure*}
    \sidecaption
    \includegraphics[width=12cm]{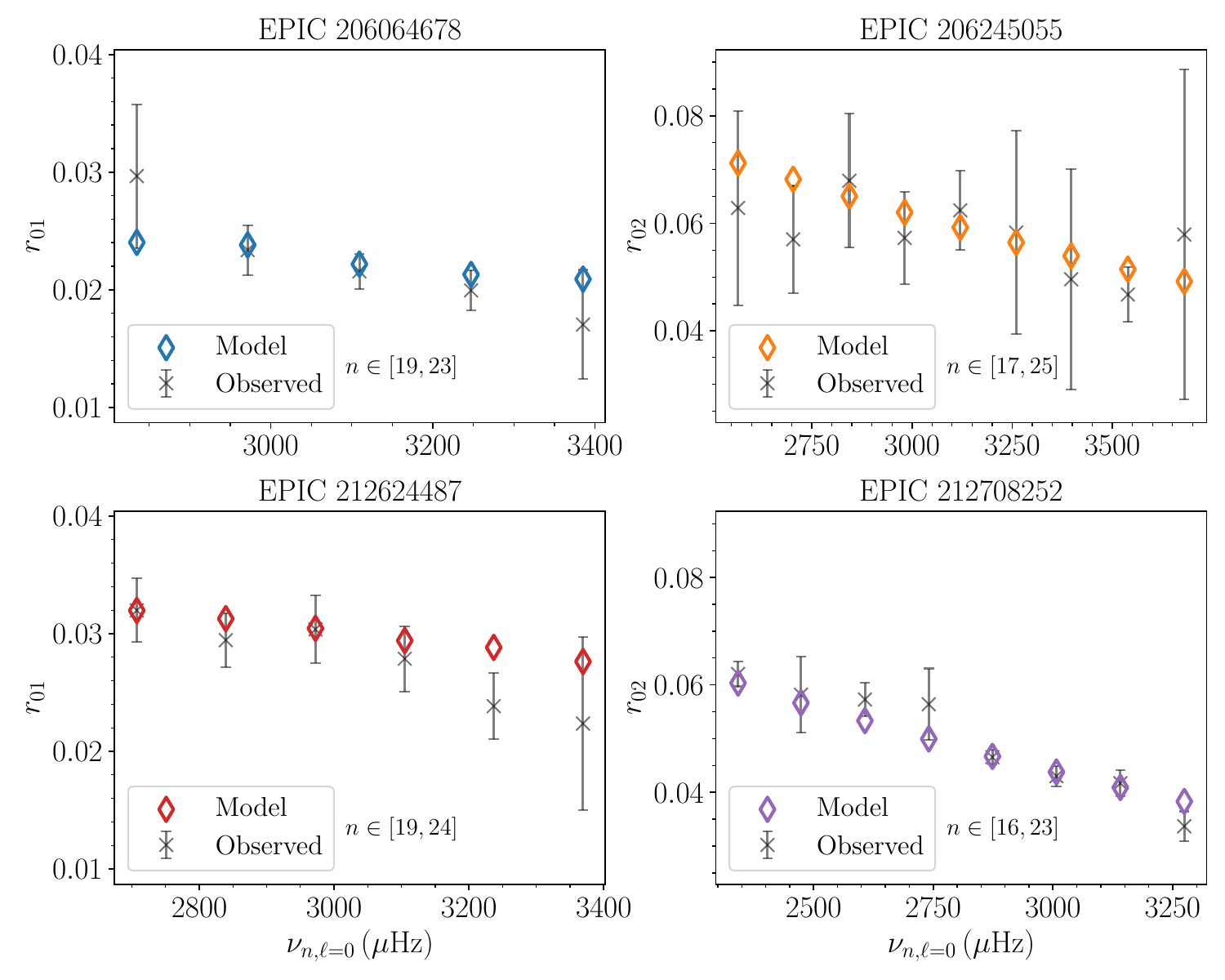}
        \caption{Frequency separation ratios \rone or \rtwo for the OPAL+NACRE II optimal models after the Levenberg-Marquardt minimisation of EPIC 206064678 (top left), EPIC 206245055 (top right), EPIC 212624487 (bottom left), and EPIC 212708252 (bottom right panel), as a function of the observed frequency. The modelled and observed ratios are shown as coloured diamonds and grey crosses, respectively. Note the different $y$-axes for the \rone (left panels) and the \rtwo (right panels).}
        \label{fig:ratios}
\end{figure*}

\begin{figure}
    \centering
    \includegraphics[width=\hsize]{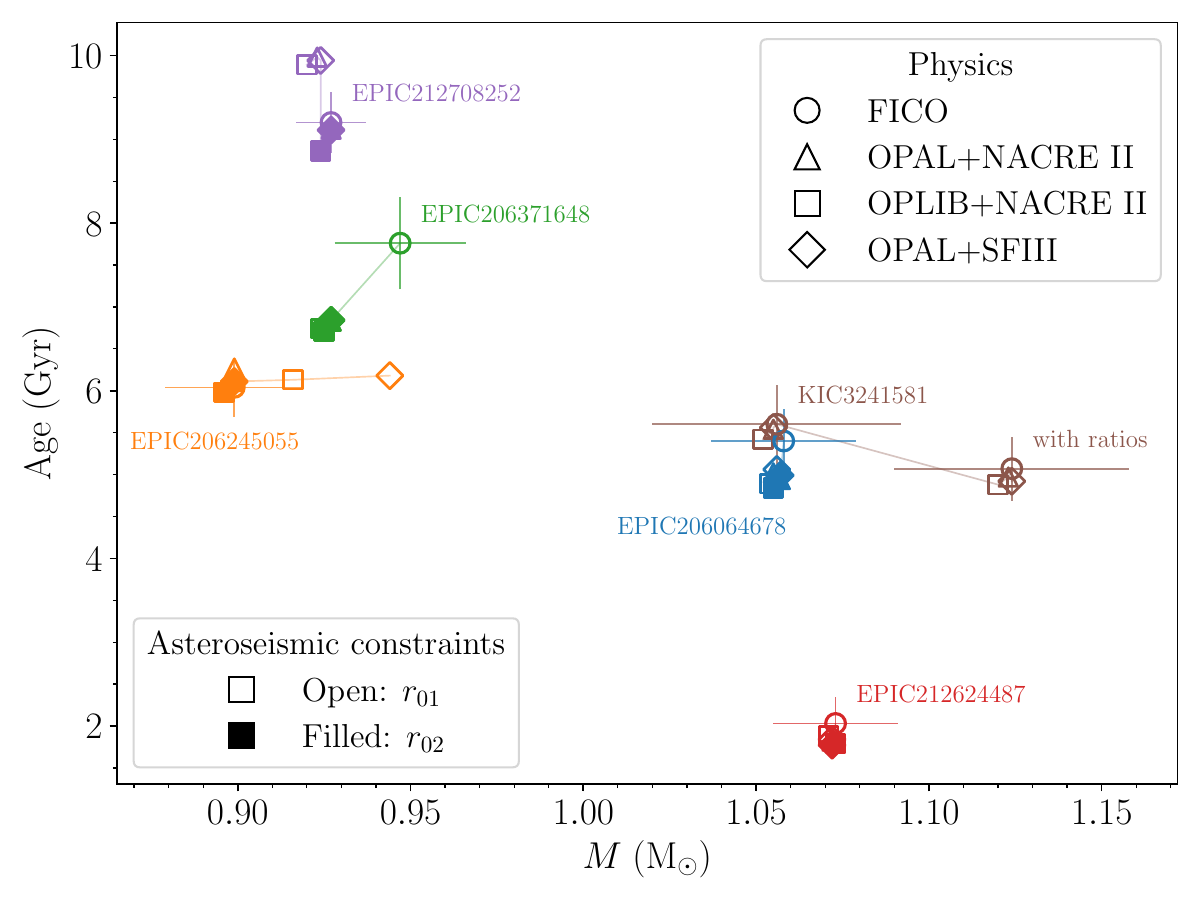}
        \caption{Asteroseismic mass-age relation illustrating the differences among the FICO procedure (circles) and the local Levenberg-Marquardt minimisations for different input physics: OPAL and NACRE II (triangles), OPLIB and NACRE II (squares), and OPAL and SFIII (diamonds). The latter have been calibrated such as to reproduce the observed lithium abundances. The open and filled symbols correspond to the minimisation using \rone and \rtwo ratios, respectively. For KIC 3241581, one set corresponds to the minimisation with frequencies only and the other one with frequencies and two ratio estimates.}
        \label{fig:massage}
\end{figure}

\begin{figure*}
    \centering
    \includegraphics[width=\hsize]{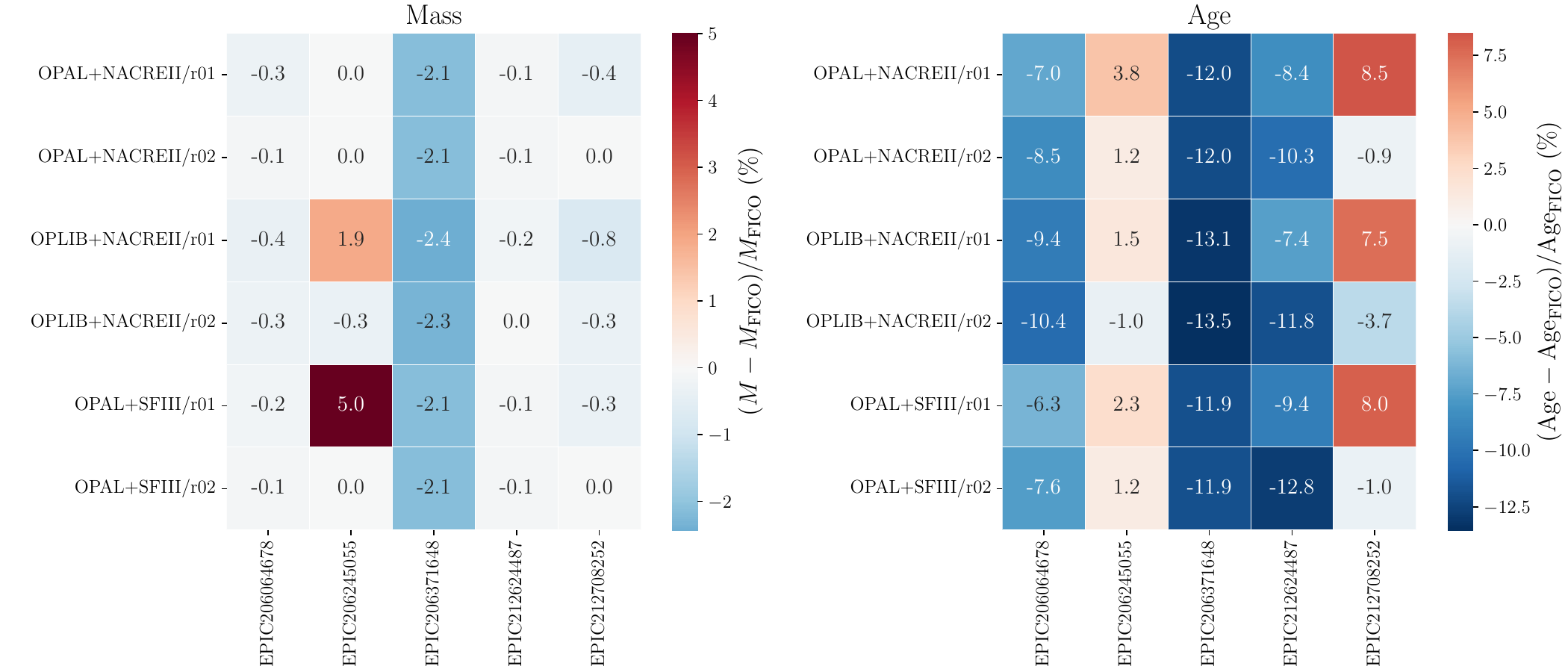}
        \caption{Heatmaps illustrating the relative variations (\%) on the mass (left) and age (right panel), between our local minimisations and the FICO results -- which we consider as the reference. Each row corresponds to a given combination of physics (OPAL+NACRE II, OPLIB+NACRE II, or OPAL+SFIII) and ratios (\rone or \rtwo); and each column corresponds to a given target.}
        \label{fig:heatmap}
\end{figure*}

\section{Non-standard processes and local minimisation}
\label{sec:localmin}

As presented in Sect. 7.3 by \citet{Garcia2026}, HERMES spectroscopy was obtained for these solar analogs -- with observations spanning $\sim 400$ days -- and allowed, in some cases, for a determination of the lithium abundance within 0.05 dex (for EPIC 206245055 and EPIC 212624487) or at least for an upper boundary limit (for the other four targets; see Table \ref{tab:obs}). Because of the lower precision of these measurements with respect to other observational constraints, such as oscillation frequencies for example, we decided not to explicitly include this information in the modelling procedure. Instead, we verify a posteriori that our optimal solution effectively reproduces the observed lithium abundance within $1\sigma$, when an actual estimate was available, or that it lies below the upper limit provided. Hence, these lithium measurements provide us with information to calibrate the amount of extra mixing required in the non-standard models in order to best reproduce the observations.

Non-standard physical processes are typically invoked in the form of either turbulent diffusion or convective boundary mixing below the envelope. We briefly discuss both of these processes and their main characteristics below.

Turbulent diffusion is modelled as in \citet{Proffitt1991}, with the following equation:
\begin{align}
    D_{\rm turb} = D_{\rm t} \left( \frac{\rho_{\rm BCZ}}{\rho} \right)^n \rm \, ,
\end{align}
where $\rho_{\rm BCZ}$ is the density at the base of the convective zone (BCZ), $\rho$ is the local density, while $D_{\rm t}$ and $n$ are the two free parameters in this formalism.

Convective boundary mixing at the base of the envelope (or convective overshooting) is also considered, using a simple convective penetration formalism \citep[see, e.g.,][]{Zahn1991}:
\begin{align}
    d_{\rm over} = \alpha_{\rm over} \, H_{\rm P} \rm \, , 
\end{align}
where $d_{\rm over}$ is the distance over which additional mixing beyond the convective boundary is acting, $\alpha_{\rm over}$ is a free parameter similar to the mixing-length parameter, and $H_{\rm P} = - \, dr/d \ln P$ is the pressure scale height.

In our study, we test the impact of using different physical ingredients within CLES on the resulting stellar parameters. Notably, we look into the influence of the opacity tables (Sect. \ref{sec:opacities}) -- OPAL or OPLIB \citep{Colgan2015} -- and of the nuclear reaction rates (Sect. \ref{sec:nuclear}) -- combination of NACRE II and SFII or Solar Fusion III \citep[SFIII,][]{Acharya2025}, to which we also add the impact of additional mixing calibrated to reproduce the observed lithium depletion. Hence, we run a local minimisation on three different model sets: OPAL+NACRE II, OPLIB+NACRE II, and OPAL+SFIII. Turbulent diffusion is always included in those non-standard models, with an efficiency of $D_{\rm t}=3500$ or higher (see discussion in Sect. \ref{sec:mixing}). The evolutionary tracks for models with updated physics are again shown on Fig. \ref{fig:hrd}, as dashed, dotted, and dash-dotted lines, respectively. Figure \ref{fig:agelithium} shows the evolution of the surface lithium abundance as a function of age for the non-standard models computed with OPAL opacities and NACRE II/SFII nuclear reaction rates. It provides a general illustration of how lithium depletes along the evolution and the agreement with the available observational constraints. Similar trends would also be seen in the cases of OPLIB+NACRE II and OPAL+SFIII.

The observational constraints we use within the local minimisations typically find themselves divided in two categories:
\begin{itemize}
    \item classical constraints: luminosity, metallicity, inverted mean density, and effective temperature;
    \item and asteroseismic constraints: \rone or \rtwo ratios following the definitions of \citet{Roxburgh2003}, sometimes supplemented or replaced by individual oscillation frequencies. 
\end{itemize}
There are also cases for which we drop a classical constraint and/or ratios because of their unavailability or unreliability (see Sect. \ref{sec:constraints}). Figure \ref{fig:ratios} shows how the observed and modelled, i.e. resulting from our local minimisation, ratios compare with one another for four different targets. For two of them, we preferred the fit obtained with \rone ratios, while for the other two, the solutions resulting from the \rtwo fit are more acceptable. \rtwo ratios are sometimes less numerous and subject to precision issues, due to their reliance on quadrupolar frequencies. Hence, our initial choice of ratios defaults preferentially to \rone ratios, unless they provide a solution in strong disagreement with FICO's. This is discussed in further detail in Sect. \ref{sec:ratios}. The optimal solutions resulting from our local minimisations are gathered in Table \ref{tab:min}.

In the remainder of this paper, we consider the FICO results (Table \ref{tab:fico}) as the standard and reference models, with OPAL opacities, a combination of NACRE II and SFII nuclear reaction rates, and with microscopic diffusion alone being taken into account. We thus compare the stellar fundamental parameters resulting from the latter to our local minimisations with non-standard physics, i.e. different opacity and/or nuclear reaction rates tables and the inclusion of turbulent diffusion, at the level of $D_{\rm t}=3500$ or higher, and sometimes convective overshooting at the base of the envelope. We discuss general deviations in mass and age between FICO and our models, as well as the respective influences due to changes in the opacities, nuclear reaction rates, and inclusion of extra-mixing in Sect. \ref{sec:results}. We provide more detailed insights into specific aspects related to the methodology (reliability of classical and/or seismic constraints, choice of frequency separation ratios) and how they impact the precise modelling of certain solar analogs in the context of PLATO, and the associated precision requirements, in Sect. \ref{sec:discussion}.

\begin{table*}[]
    \centering
    \caption{Stellar parameters determined from the local minimisation.}
    \begin{tabular}{c|c|c|c|ccccc}
    \hline\hline
         Target          & Physics          & Fit               & $\chi^2_{\rm red}$    & $M$ (\msun)  & $R$ (\rsun)  & $\tau$ (Gyr)  & $X_{\rm 0}$  & $Z_{\rm 0}$ \\
    \hline
         EPIC 206064678  & OPAL+NACRE II    & \rone             & 0.702        & 1.055        & 1.003        & 5.02          & 0.724        & 0.026       \\
         EPIC 206064678  & OPAL+NACRE II    & \rtwo             & 0.587        & 1.057        & 1.003        & 4.94          & 0.724        & 0.026       \\  
         EPIC 206064678  & OPLIB+NACRE II    & \rone             & 0.456        & 1.054        & 1.003        & 4.89          & 0.726        & 0.027       \\
         EPIC 206064678  & OPLIB+NACRE II    & \rtwo             & 0.425        & 1.055        & 1.003        & 4.84          & 0.726        & 0.027       \\
         EPIC 206064678  & OPAL+SFIII       & \rone             & 0.664        & 1.056        & 1.003        & 5.06          & 0.724        & 0.026       \\
         EPIC 206064678  & OPAL+SFIII       & \rtwo             & 0.645        & 1.057        & 1.003        & 4.99          & 0.724        & 0.026       \\  
    \hline
         EPIC 206245055  & OPAL+NACRE II    & \rone             & 1.643        & 0.899        & 0.944        & 6.27          & 0.740        & 0.006       \\
         EPIC 206245055  & OPAL+NACRE II    & \rtwo             & 0.976        & 0.899        & 0.943        & 6.11          & 0.739        & 0.006       \\
         EPIC 206245055  & OPLIB+NACRE II    & \rone             & 1.375        & 0.916        & 0.950        & 6.13          & 0.768        & 0.006       \\
         EPIC 206245055  & OPLIB+NACRE II    & \rtwo             & 1.020        & 0.899        & 0.942        & 5.98          & 0.741        & 0.006       \\
         EPIC 206245055  & OPAL+SFIII       & \rone             & 1.287        & 0.944        & 0.959        & 6.18          & 0.788        & 0.005       \\
         EPIC 206245055  & OPAL+SFIII       & \rtwo             & 0.949        & 0.899        & 0.943        & 6.11          & 0.739        & 0.006       \\   
    \hline
         EPIC 206371648  & OPAL+NACRE II    & \rone             & 5.100        & 0.927        & 0.968        & 6.83          & 0.738        & 0.010       \\
         EPIC 206371648  & OPAL+NACRE II    & \rtwo             & 9.725        & 0.927        & 0.968        & 6.83          & 0.738        & 0.010       \\
         EPIC 206371648  & OPLIB+NACRE II    & \rone             & 4.980        & 0.924        & 0.967        & 6.74          & 0.740        & 0.010       \\
         EPIC 206371648  & OPLIB+NACRE II    & \rtwo             & 12.05        & 0.925        & 0.967        & 6.71          & 0.739        & 0.010       \\
         EPIC 206371648  & OPAL+SFIII       & \rone             & 4.860        & 0.927        & 0.968        & 6.84          & 0.738        & 0.010       \\
         EPIC 206371648  & OPAL+SFIII       & \rtwo             & 9.825        & 0.927        & 0.968        & 6.84          & 0.737        & 0.010       \\
    \hline
         EPIC 212624487  & OPAL+NACRE II    & \rone             & 1.082        & 1.072        & 1.034        & 1.86          & 0.742        & 0.010       \\
         EPIC 212624487  & OPAL+NACRE II    & \rtwo             & 0.910        & 1.072        & 1.034        & 1.82          & 0.741        & 0.011       \\
         EPIC 212624487  & OPLIB+NACRE II    & \rone             & 0.890        & 1.071        & 1.034        & 1.88          & 0.749        & 0.010       \\
         EPIC 212624487  & OPLIB+NACRE II    & \rtwo             & 0.791        & 1.073        & 1.034        & 1.79          & 0.747        & 0.011       \\
         EPIC 212624487  & OPAL+SFIII       & \rone             & 1.055        & 1.072        & 1.033        & 1.84          & 0.741        & 0.010       \\
         EPIC 212624487  & OPAL+SFIII       & \rtwo             & 1.207        & 1.072        & 1.032        & 1.77          & 0.740        & 0.010       \\
    \hline
         EPIC 212708252  & OPAL+NACRE II    & \rone             & 1.222        & 0.923        & 0.980        & 9.98          & 0.737        & 0.013       \\
         EPIC 212708252  & OPAL+NACRE II    & \rtwo             & 1.113        & 0.927        & 0.981        & 9.12          & 0.735        & 0.013       \\
         EPIC 212708252  & OPLIB+NACRE II    & \rone             & 1.275        & 0.920        & 0.979        & 9.89          & 0.742        & 0.013       \\
         EPIC 212708252  & OPLIB+NACRE II    & \rtwo             & 1.306        & 0.924        & 0.980        & 8.86          & 0.737        & 0.013       \\
         EPIC 212708252  & OPAL+SFIII       & \rone             & 1.312        & 0.924        & 0.980        & 9.94          & 0.738        & 0.013       \\
         EPIC 212708252  & OPAL+SFIII       & \rtwo             & 1.094        & 0.927        & 0.981        & 9.11          & 0.735        & 0.013       \\
    \hline
         KIC 3241581     & OPAL+NACRE II    & \nunl             & 1.347        & 1.055        & 1.088        & 5.54          & 0.688        & 0.027       \\    
         KIC 3241581     & OPAL+NACRE II    & \nunl+\rone+\rtwo & 4.460        & 1.123        & 1.110        & 4.97          & 0.725        & 0.026       \\    
         KIC 3241581     & OPLIB+NACRE II    & \nunl             & 1.803        & 1.052        & 1.087        & 5.42          & 0.690        & 0.027       \\    
         KIC 3241581     & OPLIB+NACRE II    & \nunl+\rone+\rtwo & 4.740        & 1.120        & 1.109        & 4.88          & 0.727        & 0.027       \\    
         KIC 3241581     & OPAL+SFIII       & \nunl             & 1.357        & 1.055        & 1.088        & 5.56          & 0.689        & 0.027       \\    
         KIC 3241581     & OPAL+SFIII       & \nunl+\rone+\rtwo & 4.460        & 1.124        & 1.110        & 4.92          & 0.725        & 0.026       \\ 
    \hline
    \end{tabular}
    \tablefoot{The efficiency of macroscopic transport was calibrated on lithium abundances.}
    \label{tab:min}
\end{table*}

\section{Results}
\label{sec:results}

Figure \ref{fig:massage} shows the asteroseismic mass-age relation for our six seismic solar analogs. We consider the mass and age of the FICO model as the reference values, and quantify the deviations from those estimates due to changes in physical ingredients and the inclusion of additional mixing processes. In Sect. \ref{sec:mass-age}, we look carefully into each of these stars and how a change in the physics and/or in the constraints used can lead to systematic effects on the estimated masses and ages. Sections \ref{sec:opacities}, \ref{sec:nuclear}, and \ref{sec:mixing} relate the physical modifications considered in this work (opacities, nuclear reaction rates, and extra-mixing) with changes in the central temperature and surface chemical composition occurring along the evolution.

\subsection{Impact of non-standard physics on the determination of stellar masses and ages}
\label{sec:mass-age}

For EPIC 206064678, the variations in mass are below 0.4\%, and those in age below 10.5\%. In the case of EPIC 206245055, there is no variation in mass except in the case of \rone ratios for OPLIB+NACRE II ($\sim 2\%$) and OPAL+SFIII (5\%), which are problematic cases (very low $Y_{\rm 0}$) which we discuss further in Sect. \ref{sec:ratios}. Age-wise, deviations lie below 4\% and the most compatible ones with FICO are obtained via the minimisation of \rtwo ratios. For EPIC 206371648, the variations in mass are lower than 2.5\% and those in age all higher than 11\% (highest one for OPLIB+NACRE II with 13\%), which might be due to a combination of the unreliable luminosity estimate that we decided not to include in the local minimisations and of the poor quality of both \rone and \rtwo ratios (see Sect. \ref{sec:constraints}). For EPIC 212624487, the deviations in mass are below 0.1\%, and in age below 9.5\% (\rone) or 13\% (\rtwo). For EPIC 212708252, we observe discrepancies in mass below 0.8\% and in age of the order of 8\% using \rone ratios, and below 4\% using \rtwo ratios with OPLIB+NACRE II or even lower for the other two cases (below 1\%). For KIC 3241581, the star with the lowest seismic quality in our sample, if we were to consider the two sets (reduced and full sets of frequencies) independently, the variations in mass remain below 0.4\%, and those in age below 4\%. But, most importantly, the FICO estimates (and so the minimisation results as well) between the two sets differ by 6\% in mass and 10\% in age. These relative variations in mass and age are portrayed with heatmaps in Fig. \ref{fig:heatmap}. Since KIC 3241581 is a particular case for which individual oscillation frequencies were used, we deemed it not relevant to be included in the heatmaps. Generally speaking, we find that the highest deviations usually occur in the case of OPLIB+NACRE II. This is to be expected as it is known that the OPAL and OPLIB opacity tables themselves do not converge in terms of their raw values \citep[see, e.g.,][for the solar case]{Buldgen2025b}. Hence, one may expect even larger errors than the differences that we map here, which result from additional modifications to the opacity tables in the future, which will impact the inference of stellar fundamental parameters.

\subsection{Opacities}
\label{sec:opacities}

Using OPLIB opacities leads to a small shift in $T_{\rm c}$ consistent with the one observed with SFIII reaction rates for EPIC 206245055 (see left panel in Fig. \ref{fig:EPIC206245055}). But in the case of EPIC 212624487, on the left panel of Fig. \ref{fig:EPIC212624487}, the decrease in $T_{\rm c}$ is much more significant. The evolutionary tracks using OPLIB opacities are a bit less luminous than those with OPAL ones (see Fig. \ref{fig:hrd}). On the right panels of Figs. \ref{fig:EPIC206245055} and \ref{fig:EPIC212624487}, one can also observe that the surface helium abundance is lower when using OPLIB opacities. As discussed in \citet{Buldgen2019} in the context of the solar modelling problem, OPLIB tables lead to significantly lower opacity values, with respect to OPAL, in the core. Thus, in order to compensate for the decrease in opacity and still reproduce an observed luminosity at a given age, the initial hydrogen abundance increases, and the helium abundance decreases.

\subsection{Nuclear reaction rates}
\label{sec:nuclear}

The use of SFIII reaction rates leads to slightly lower $T_{\rm c}$ with respect to NACRE II, as can be seen on the left panels of Figs. \ref{fig:EPIC206245055} and \ref{fig:EPIC212624487}. Based on Fig. \ref{fig:hrd}, evolutionary tracks with NACRE II and SFIII reaction rates seem to coincide with one another. The SFIII reaction rates for the pp chain have a higher efficiency compared to the NACRE II combinations (with SFII). The pp chain's higher efficiency thus requires to compensate by lowering $X_{\rm c}$ and $T_{\rm c}$ \citep[see, e.g.,][for the solar case]{Sandron2026}.

\subsection{Extra-mixing processes: turbulent diffusion and convective overshooting}
\label{sec:mixing}

In the FICO reference models, only microscopic diffusion is considered. While, in our minimisation models, we find that the inclusion of turbulent diffusion with at least an efficiency of $D_{\rm t} = 3500$ and $n=4$, typically what one would expect for the Sun  \citep{Buldgen2025a}, is required to reproduce the lithium and beryllium observations. Although this was not enough for some stars, and we had to increase $D_{\rm t}$ to 6000-9000 for EPIC 206245055, and to 7500-9000 for EPIC 206371648. In both cases, the highest $D_{\rm t}$ value is required in the case of OPLIB+NACRE II. A more complex case was that of EPIC 212624487 where we had to combine both a higher turbulent diffusion efficiency ($D_{\rm t}=10000$ and $n=2$) and the inclusion of convective overshooting ($\alpha_{\rm over}=0.15$).

\begin{figure*}
    \sidecaption
    \includegraphics[width=6cm]{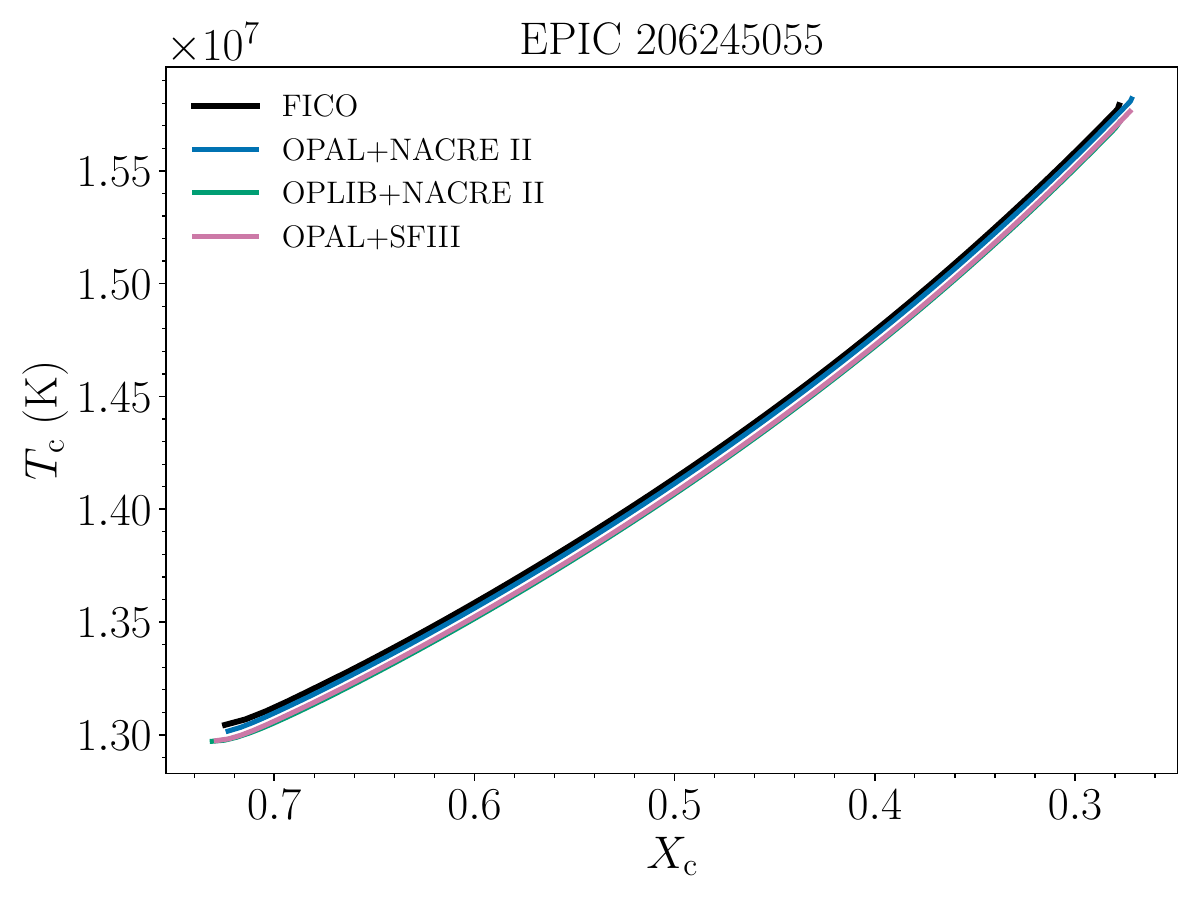}
    \includegraphics[width=6cm]{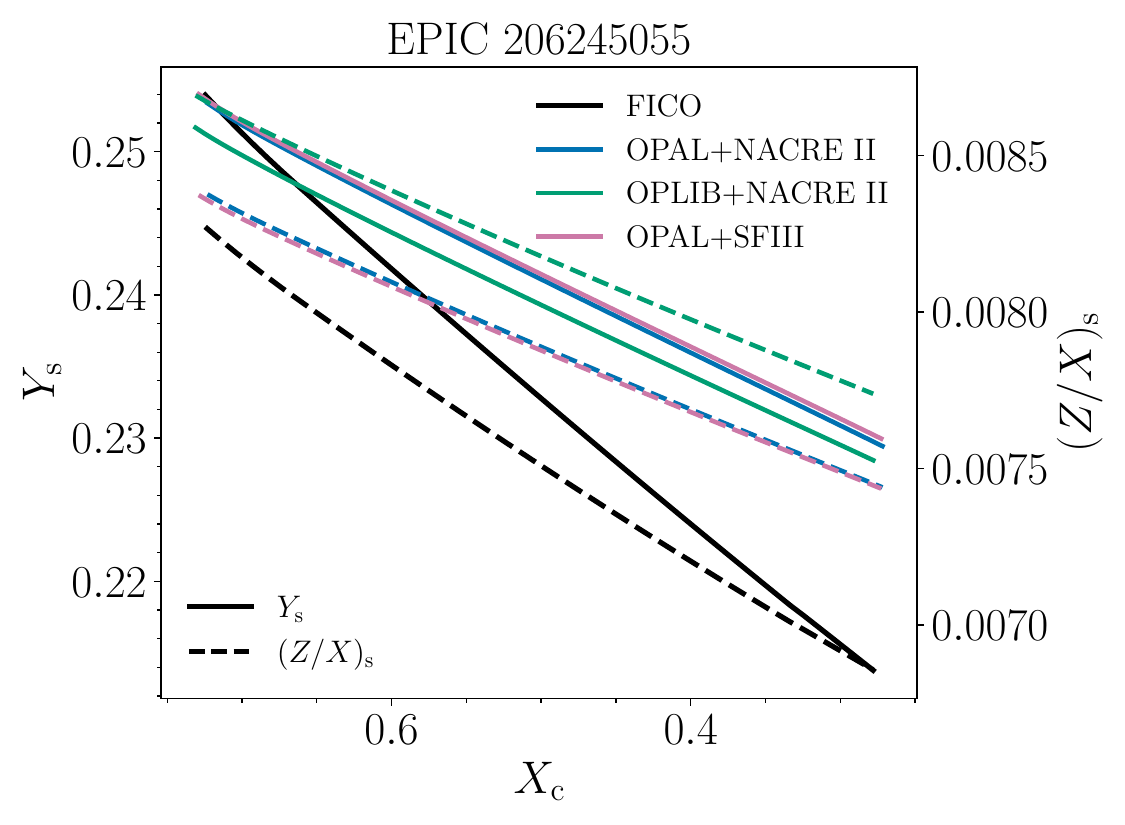}
    \caption{Evolution of the central temperature $T_{\rm c}$ (left) and of the surface chemical composition (right panel) as a function of the central hydrogen abundance $X_{\rm c}$ for EPIC 206245055. Black, blue, green, and pink-coloured lines correspond to FICO, OPAL+NACRE II, OPLIB+NACRE II, and OPAL+SFIII, respectively. On the right panel, solid and dashed lines show $Y_{\rm s}$ and $(Z/X)_{\rm s}$, respectively.}
    \label{fig:EPIC206245055}
\end{figure*}

\begin{figure*}
    \sidecaption
    \includegraphics[width=6cm]{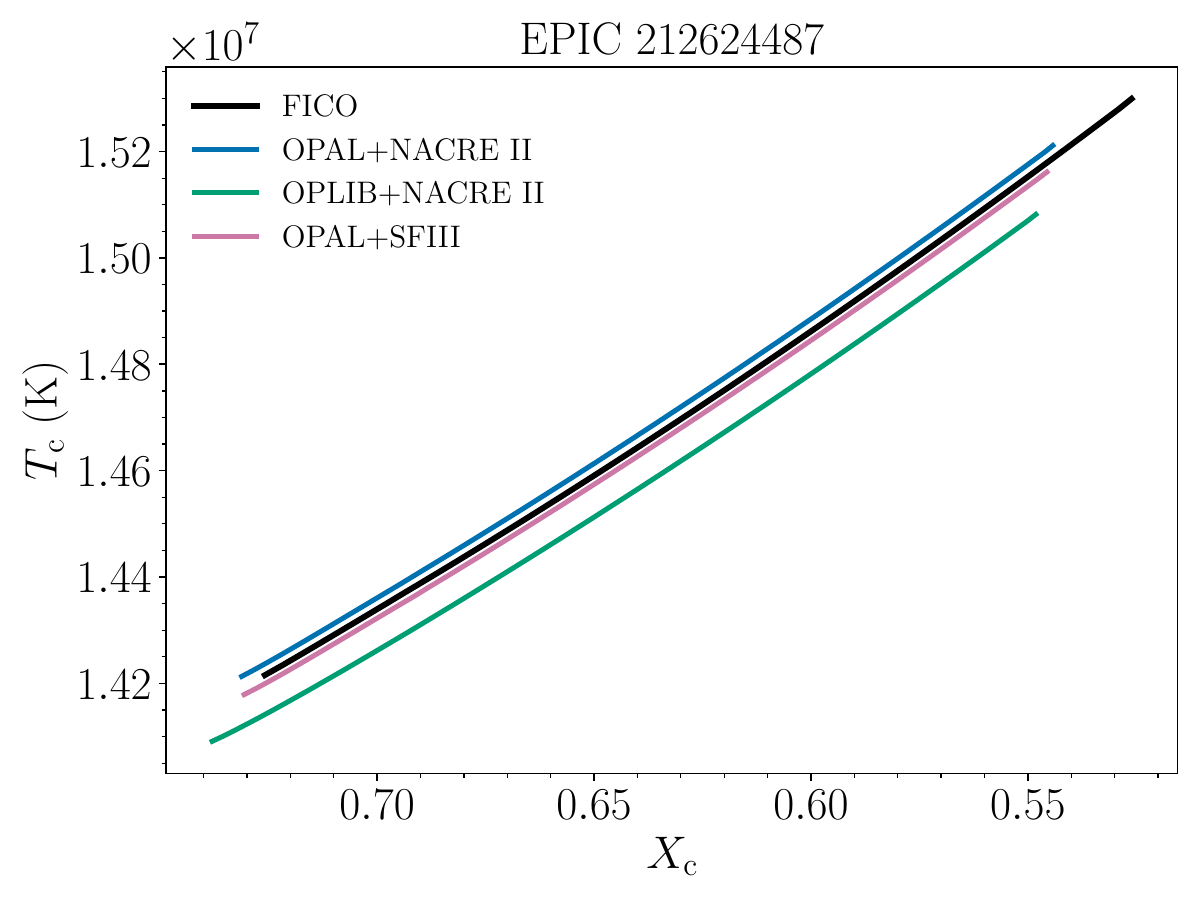}
    \includegraphics[width=6cm]{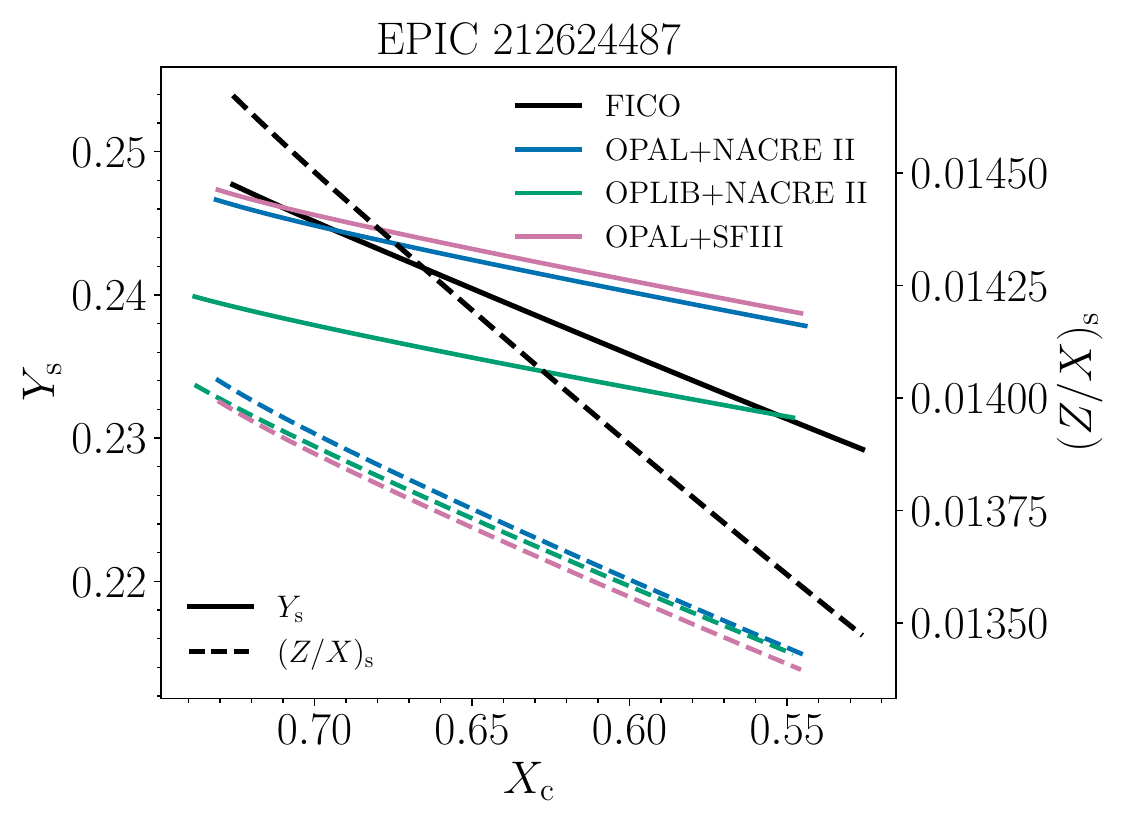}
    \caption{Same as Fig. \ref{fig:EPIC206245055} for EPIC 212624487.}
    \label{fig:EPIC212624487}
\end{figure*}

In Figs. \ref{fig:EPIC206245055} and \ref{fig:EPIC212624487}, the plots on the right show the evolution of $Y_{\rm s}$ and $(Z/X)_{\rm s}$ as a function of the central hydrogen abundance, $X_{\rm c}$, for EPIC 206245055 and EPIC 212624487. These show that the impact of macroscopic mixing is not just on the "trace" elements such as Li and Be but also on $Y$. The decrease in $Y_{\rm s}$ and $(Z/X)_{\rm s}$ is significantly steeper in the case of FICO models, where only microscopic diffusion was considered. The inclusion of turbulent diffusion leads to milder slopes for both stars. In the case of EPIC 212624487, the addition of convective overshooting below the envelope ($\alpha_{\rm ov}=0.15$) does not seem to have an impact. What is interesting to note also is that $Y_{\rm s}$ is much lower when using OPLIB+NACRE II. The effect is also present for EPIC 206245055, although not as significant.

The different $D_{\rm t}$ values could indicate a different efficiency of the chemical mixture during the lifetime of each star. We did not consider the case with envelope overshooting alone, which would just lead to a plateau \citep[see Fig. 5 in][]{Thevenin2017}. The expected temporal evolution goes in favour of turbulent diffusion \citep[see, e.g.,][]{Kunitomo2025}. Besides, envelope overshooting slightly pushes the base of the convective zone, with a small influence on microscopic diffusion, but there is no impact on the chemical evolution or nuclear reaction rates \citep[see, e.g.,][for the solar case]{Buldgen2025a}. Nevertheless, one would need more precise measurements of lithium and beryllium in order to disentangle the different contributions, due to turbulent diffusion or to convective overshooting \citep[see, e.g.,][]{Deal2015,Amarsi2024}.

\section{Methodology discussion and consequences for PLATO}
\label{sec:discussion}

The current study lays the groundwork for future detailed asteroseismic modelling of solar analogs in the context of PLATO. We discuss below methodological aspects related to our modelling procedure, regarding the quality of the constraints used and the available choice among frequency separation ratios, in Sects. \ref{sec:constraints} and \ref{sec:ratios}, respectively. Section \ref{sec:plato} discusses our results in the light of the PLATO precision requirements for stellar parameters.

\subsection{Unreliable classical or seismic constraints}
\label{sec:constraints}

In some cases, the uncertainty related to either classical and/or seismic constraints does not allow us to reliably determine stellar fundamental parameters.

\begin{figure}
    \centering
    \includegraphics[width=\hsize]{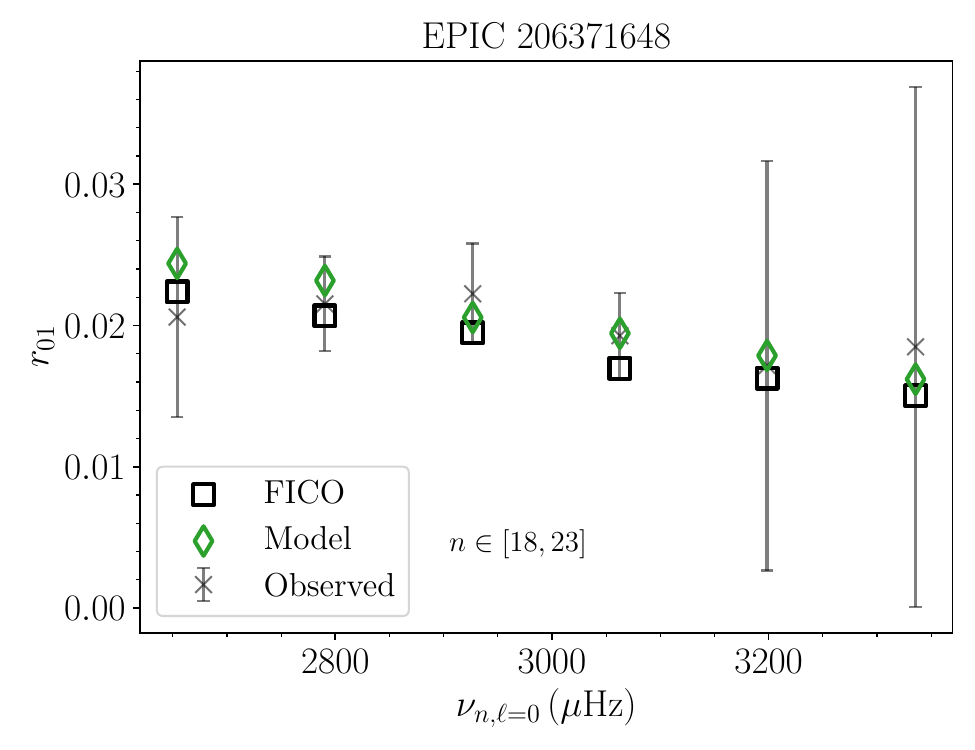}
        \caption{Frequency separation ratios \rone for EPIC 206371648 as a function of the observed frequency. The FICO, modelled (OPAL+NACRE II minimisation) and observed ratios are shown as black squares, coloured diamonds and grey crosses, respectively.}
        \label{fig:EPIC206371648}
\end{figure}

For EPIC 206371648, the third step of the FICO method (with ratios) did not converge, and so only the results from the modelling of individual frequencies were provided. We were still able to calculate six consecutive \rone ratios with $n=[18,23]$ and five non-consecutive \rtwo ratios with $n=\{18,19,21,23,24\}$. However, we note that both \rone and \rtwo ratios have large uncertainties, although our predictions do seem to reproduce them well (see Fig. \ref{fig:EPIC206371648} for \rone ratios). We also added the lowest radial-order frequency ($n=17$), in order to avoid a shift in the échelle diagram. Contrarily to the other targets, we removed the luminosity from the classical constraints because the high \texttt{RUWE} = 9.1 and the 6-parameter astrometric solution could suggest a low astrometric quality of the \textit{Gaia} parameters. In addition, luminosity estimates differ by close to $4\sigma$ depending on whether one uses the $K_s$ or $G$ band for the luminosity computation. For this reason, we preferred to discard the luminosity constraint within the fit. As things were, we had obtained a solution with a near-solar metallicity and a helium abundance significantly lower than the primordial value, which was not acceptable. Hence, we also decreased $(Z/X)_{\rm 0}$ and used a helium enrichment law: $Y=Y_{\rm P} + \frac{\Delta Y}{\Delta Z} Z$, with $\Delta Y/\Delta Z = 1.1$ and $Y_{\rm P}=0.2485$. The solar values for OPAL and NACRE II are X$_{\rm \odot}=0.72$,  Z$_{\rm \odot}=0.015$, and Y$_{\rm \odot}=0.265$. Despite these adjustments, we are able to better recover the metallicity \feh ($1.5\sigma$ away from the observed value) but the minimisation fits fail to reproduce the effective temperature \teff ($4.6\sigma$ away from the observed value). Hence, for this star, both the classical (luminosity) and seismic constraints (\rone and \rtwo ratios) are at fault and prevent an accurate modelling of its stellar parameters. Further analysis would be required to quantify potential discrepancies induced by the minimisation procedure itself.

In the same vein for KIC 3241581, only the results from the modelling step with individual frequencies were provided by the FICO method. Two different solutions are given, either based on a reduced set of frequencies where only those determined in common by two independent asteroseismic pipelines (\texttt{apollinaire}, \citealt{Breton2022}; and \texttt{PBjam}, \citealt{Nielsen2021,Nielsen2025}) were retained and a full set where all frequencies were kept. Hence, we run two different sets of minimisation: using the three lowest frequencies for $\ell=\{0,1,2\}$, with $n=21$, 20, and 19, respectively, in the first case; and using the three lowest frequencies for $\ell=\{0,1,2\}$, with $n=16$, 16, and 18, respectively, as well as \rone ($n=20$) and \rtwo ($n=20$), in the second case. The significant differences resulting from those two solutions make it difficult to draw reliable conclusions regarding the mass and age of this star.

\subsection{\rone vs \rtwo ratios}
\label{sec:ratios}

Given that \rtwo ratios are sometimes less numerous and less precise, as they involve quadrupolar frequencies that exhibit here larger uncertainties, we mainly focused on interpreting the results based on our \rone minimisations. There are however two cases where the fits with \rone ratios turned out to be problematic. In the case of EPIC 206245055, they do not allow us to retrieve a reliable estimate of the mass, and there are very significant differences depending on the physics used. Especially for OPLIB+NACRE II and OPAL+SFIII, they lead to non-physical solutions where $Y_{\rm 0}$ is low to the point that it lies below the primordial helium abundance. As for EPIC 212708252, the \rone minimisation results significantly lie above the FICO solution in terms of age. 

\begin{figure*}
    \sidecaption
    \includegraphics[width=12cm]{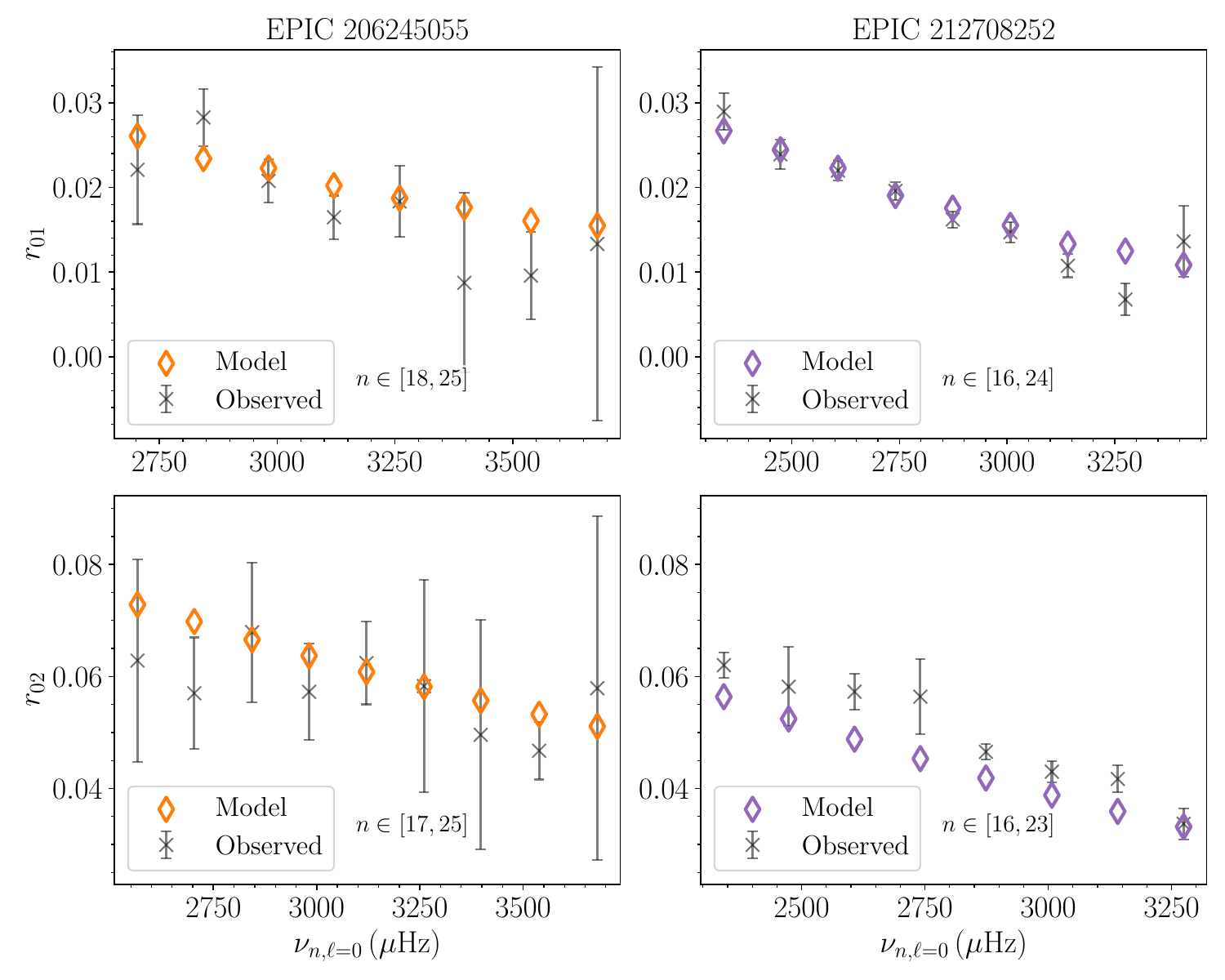}
        \caption{Frequency separation ratios \rone based on the OPAL+SFIII fit with \rtwo (top) and \rtwo based on the OPAL+SFIII fit with \rone for EPIC 206245055 (left) and EPIC 212708252 (right), as a function of the observed frequency. The modelled and observed ratios are shown as coloured diamonds and grey crosses, respectively. Note the different $y$-axes for the \rone (top panels) and the \rtwo (bottom panels).}
        \label{fig:r01tor02}
\end{figure*}

For these two targets, we also have a look at the predicted \rone ratios based on the \rtwo minimisation, and the predicted \rtwo ratios based on the \rone minimisation. These are shown and compared to the observed ratios on Fig. \ref{fig:r01tor02}. In the case of EPIC 206245055 (left panels), both predicted \rone and \rtwo ratios stay mostly within the observed error bars, but the latter are also quite large. As for EPIC 212708252 (right panels), the observed uncertainties are smaller and one can directly tell that the predicted \rtwo ratios resulting from the \rone fit do not match the observations. For this reason, we decided to focus only on the \rtwo fit.

\subsection{PLATO requirements}
\label{sec:plato}

The requirements for the PLATO mission are such that one expects precisions of 15\%, 2\%, and 10\% for the stellar mass, radius, and age, respectively. However, to be fully consistent, one should aim to estimate the mass at a precision level of 3\% in order to reach a 10\% precision in age. Regarding the mass, the deviations that we measure never go beyond 2.5\% (except for the problematic case of EPIC 206245055 with \rone ratios). Hence, the 3\% precision requirement is always fulfilled. The precision requirement for the stellar radius is similarly achieved, as deviations remain below 1\% (see Fig. \ref{fig:heatmap_r}). As for the age, the picture becomes a bit more complicated. For one third of the cases, the deviations from the FICO age are above 10\% but still below 14\%. For the other two thirds, the precision requirement of 10\% in age is achievable, and two particularly good cases are the ones of EPIC 206245055 and EPIC 212708252 where the fit of \rtwo ratios leads to deviations as low as 4\%.

\section{Conclusions}
\label{sec:conclusions}

In this paper, we have carried out a detailed modelling analysis for the five K2 and one \textit{Kepler} seismic solar analogs studied in \citet{Garcia2026}, for which constraints on the surface lithium abundance were available thanks to HERMES spectroscopy data. In Sect. \ref{sec:reference}, we briefly present the FICO method \citep{Betrisey2023,Betrisey2026} and the reference results that it leads to, in terms of mass, radius, and age. In Sect. \ref{sec:localmin}, we present our approach -- which consists in building upon the FICO results to perform Levenberg-Marquardt minimisations with non-standard physical processes. These include both changes in the opacity (OPAL or OPLIB) and nuclear reaction rates tables (NACRE II or SFIII) used, but also the addition of extra mixing in the form of turbulent diffusion and/or convective boundary mixing below the envelope in order to reproduce the observed lithium depletion. Both classical, i.e. luminosity, metallicity, mean density, and effective temperature, and asteroseismic constraints, i.e. \rone or \rtwo ratios (individual frequencies in the case of KIC 3241581), take part in the modelling procedure. 

Our results are presented in Sect. \ref{sec:results}, where we quantify the deviations in mass and age between our minimisations and the FICO results, which we consider as the reference. In terms of mass, the deviations always stay below 2.5\%, except for the case of EPIC 206245055 with OPAL+SFIII physics and \rone ratios (5\%). However, it still remains within the 15\% requirement for PLATO and we have also found that the \rone minimisation fits for this star were not reliable enough. In terms of age, the picture becomes a bit more complicated. For EPIC 206245055 and EPIC 212708252, we decided to focus on the \rtwo minimisation results and the deviations in age lie around $\sim 4\%$ at most, hence well below the 10\% requirement. For EPIC 206064678 and EPIC 212624487, the relative variations are either below or just above 10\%, depending on the physics and ratios used. Minimisation fits based on \rone ratios seem to consistently lead to variations below 10\% for these two targets. EPIC 206371648 is the most problematic case, for which all age predictions deviate by more than 11\% with respect to FICO. But this is also a particular case, where we know that both classical and asteroseismic constraints do not allow for an accurate modelling of the stellar parameters. This thus demonstrates the importance of detailed seismology and of non-seismic parameters in the context of PLATO. Sections \ref{sec:opacities} and \ref{sec:nuclear} illustrate how a change in the opacity or nuclear reaction rates tables leads to shifts in the central temperature, $T_{\rm c}$, consistent with what has been previously modelled in the solar case \citep[see, e.g.,][]{Buldgen2019,Sandron2026}. As opposed to the FICO reference models, where only microscopic diffusion was taken into account, our modelling procedure always includes turbulent diffusion to some extent and, sometimes, convective boundary mixing below the envelope. These aspects and, in particular, the differences that they trigger on the evolution of the surface chemical composition ($Y_{\rm s}$ and $(Z/X)_{\rm s}$) are covered in Sect. \ref{sec:mixing}.

In the study by \citet{Garcia2026}, a multi-pipeline global minimisation had been carried out to estimate the stellar parameters for these stars, based on six different stellar evolution codes and model optimisation methods. Their conclusions regarding the estimation of stellar parameters converge with ours, in the sense that we are very well able to retrieve the stellar mass and radius within 3\% and 1-2\%, respectively. As for stellar ages, they are determined with a precision spanning a range from $\sim$ 5\% to 25\% -- the highest deviations, due to using different modelling methods (which may also involve different physics), pertaining to EPIC 206371648 and KIC 3241581, which we know have issues or higher uncertainties in their classical and/or asteroseismic constraints.

Section \ref{sec:discussion} focuses on methodological questions related to the modelling procedure used in this paper. The questions related to the reliability of the classical and seismic constraints involved in the minimisation procedure are tackled in Sects. \ref{sec:constraints} and \ref{sec:ratios}. The former discusses the specific cases of EPIC 206371648 and KIC 3241581, where the poor quality of the asteroseismic constraints (and also of the luminosity for EPIC 206371648) prevents us from drawing conclusions regarding the mass and age of these targets. The latter section focuses specifically on the use of either \rone or \rtwo ratios within the minimisation, and how one may obtain erroneous results when blindly using \rone ratios alone. In the framework of our study, this impacts EPIC 206245055 and EPIC 212708252. In Sect. \ref{sec:plato}, we discuss our results more specifically in the context of the PLATO precision requirements for the stellar mass, radius, and age.

These K2 and \textit{Kepler} seismic solar analogs constitute an exquisite sample to test detailed asteroseismic modelling and the influence of different physics, such as opacities, nuclear reaction rates, and also extra-mixing processes -- thanks to the HERMES observational constraints on their lithium depletion. In this study, we are able to quantify the deviations in mass and age induced by changes in the model physical ingredients, and to illustrate how they fit within the bigger picture of the PLATO requirements, i.e. 15\% in mass and 10\% in age. As already discussed for the Sun, there is no clear indication as to why one should rather use OPAL or OPLIB opacities (and, similarly, NACRE II or SFIII nuclear reaction rates), and there is room for improvement in order to better understand the consequences resulting from these different physics. Further work is also necessary to disentangle the cases where the use of \rone or \rtwo becomes more appropriate, and we stress that one should not blindly trust one or the other as this can lead to problematic solutions. Finally, additional physical ingredients worth exploring relate, among others, to the proper modelling of microscopic transport processes, as previously done, e.g. in \citet{Nsamba2018,Moedas2024}.

\begin{acknowledgements}
      We wish to thank the reviewer whose comments helped clarify and improve the paper. SK thanks the Belgian Federal Science Policy Office (BELSPO) for the financial support in the framework of the PRODEX Programme of the European Space Agency (ESA) under contract number 4000150852. GB acknowledges funding from the Fonds National de la Recherche Scientifique (FNRS) as a postdoctoral researcher. JB acknowledges funding from the SNF Postdoc.Mobility grants no. 222217 (Impact of magnetic activity on the characterization of FGKM main-sequence host-stars) and no. 239064 (Asteroseismology of Planetary Accretion on Low-Mass Host Stars). RAG acknowledges financial support from the Centre national d’études spatiales (CNES), France (ROR: https://ror.org/04h1h0y33), within the framework of the PLATO and GOLF/SoHO space missions. SM acknowledges support from the Spanish Ministry of Science and Innovation with the grant no. PID2023-149439NB-C41 and from the Agencia Estatal de Investigación (AEI) through the Severo Ochoa Centre of Excellence accreditation awarded to the Instituto de Astrofísica de Canarias, grant CEX2025-001609-S, funded by MICIU/AEI/10.13039/501100011033.
\end{acknowledgements}

\bibliographystyle{aa}
\bibliography{references}

\begin{appendix}
\section{Relative variations on the stellar radius}
\begin{figure}[h!]
    \centering
    \includegraphics[width=\hsize]{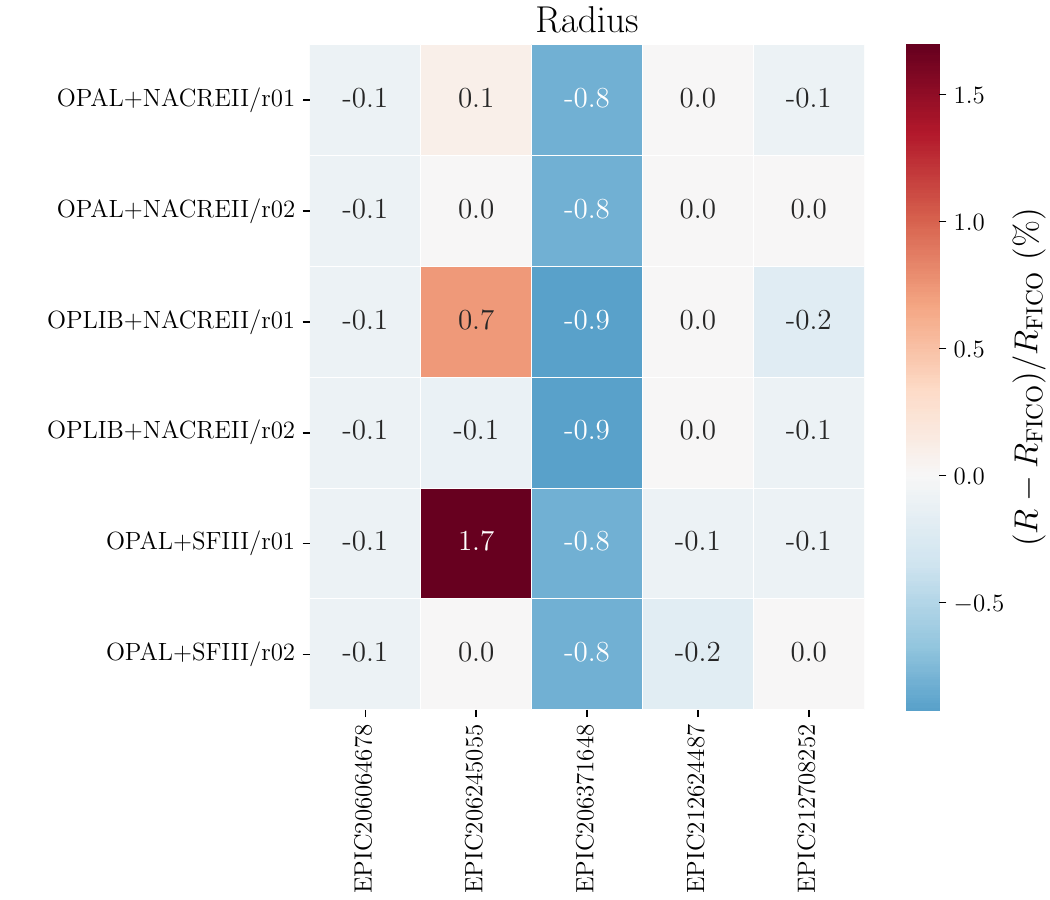}
        \caption{Same as Fig. \ref{fig:heatmap} for the radius.}
        \label{fig:heatmap_r}
\end{figure}

Figure \ref{fig:heatmap_r} portrays the relative variations (\%) on the radius between our local minimisations and the FICO results, taken as the reference.

\end{appendix}
\end{document}